\documentclass[10pt]{article}
\usepackage[letterpaper,margin=1in]{geometry}
\usepackage{amsmath,amssymb,mathrsfs}            %
\usepackage{graphicx}                            %
\usepackage{stmaryrd}
\usepackage{setspace}
\usepackage{alltt}
\usepackage{amsthm}

\usepackage[linesnumbered,ruled]{algorithm2e} %
\SetKwFunction{SAT}{SAT}
\SetKwFunction{cnf}{CNF}
\SetKwFunction{solve}{Solve}
\SetKwFunction{refine}{Refine}
\SetKwFunction{prenex}{Prenex}
\SetKwData{out}{outc}%
\SetKwData{nul}{NULL}%
\SetKwData{false}{false}\SetKwData{true}{true}
\SetKwInOut{Input}{input}
\SetKwInOut{Output}{output}
\SetKwInOut{Prec}{precondition}
\DontPrintSemicolon%
\SetKw{Func}{Function}
\SetKwFunction{addvariable}{addVars}
\SetKwFunction{addformula}{addFla}

\newcommand{\qt}[1]{\textrm{``}#1\textrm{''}}

\newcommand{\mathematics}[1]{\hbox{${#1}$}}
\newcommand{\comprehension}[2]{\ensuremath{\left\{ {#1} \;|\; {#2}\right\}}}

\DeclareMathOperator*{\bset}{\mathcal{B}}
\DeclareMathOperator*{\M}{\mathcal{M}}

\DeclareMathOperator*{\var}{{\sf var}}
\newtheorem{observation}{Observation}

\def\miko{Mikol\'a\v{s} Janota}
\def\jpms{Joao Marques-Silva}
\def\ed{Edmund Clarke}
\def\will{William Klieber}
\def\theTitle{Solving QBF with Counterexample Guided Refinement}
\def\theTitleBroken{Solving QBF with\\ Counterexample Guided Refinement}
\def\rareqs{\xspace{\rm RAReQS}\xspace}
\def\qube{\xspace{\sf QuBE7.2}\xspace}
\def\quantor{\xspace{\sf Quantor}\xspace}
\def\nenofex{\xspace{\sf Nenofex}\xspace}
\def\GQ{\xspace{\sf GhostQ}\xspace}
\def\GQc{\xspace{\sf GhostQ-CEGAR}\xspace}
\def\bloqqer{\xspace{\sf bloqqer}\xspace}
\def\QUBOS{\xspace{\sf QUBOS}\xspace}
\usepackage[pdftex]{color}
\definecolor{citeblue}{rgb}{0.1,0,.4}
\definecolor{refcolor}{rgb}{0,0,0.4}
\definecolor{midgreen}{RGB}{0,150,0}
\definecolor{darkgreen}{RGB}{0,128,0}
\definecolor{darkblue}{RGB}{0,0,128}
\definecolor{darkred}{RGB}{192,0,0}
\usepackage[pdftex%
,colorlinks=true%
,bookmarks=true%
,linkcolor=citeblue%
,citecolor=citeblue%
,urlcolor=blue%
,plainpages=false]{hyperref}
\hypersetup{%
pdfauthor={{\miko, \will, \jpms, \ed}},
pdftitle={{\theTitle}}
}

\makeatletter
\newcounter{@inst}
\newcounter{@auth}

\def\clearheadinfo{%
  \gdef\@author{No Author Given}%
  \gdef\@title{No Title Given}%
  \gdef\@institute{No Institute Given}%
}
\def\institute#1{\gdef\@institute{#1}}
\def\institutename{\par
 \begingroup
 \parskip=\z@ \parindent=\z@
 \setcounter{@inst}{1}%
 \def\and{\par\stepcounter{@inst}%
   \noindent$^{\the@inst}$\enspace\ignorespaces}%
 \noindent$^{1}$\enspace\ignorespaces
 \@institute\par
 \endgroup}
\def\inst#1{\unskip$^{#1}$}
\clearheadinfo

\renewcommand\maketitle{%
  \newpage
  \begingroup
    \parindent=\z@
    \renewcommand\thefootnote{\@fnsymbol\c@footnote}%
    \@ifundefined{Hy@colorlink}{}{\def\Hy@colorlink##1{\begingroup}}%
    \global\@topnum\z@
    \@maketitle
    \thispagestyle{plain}\@thanks
  \endgroup
  \setcounter{footnote}{0}%
  \clearheadinfo}

\def\@maketitle{%
  \newpage
  \def\lastand{\ifnum\value{@inst}=2\relax\unskip\ and\ \else\unskip, and\ \fi}%
  \def\and{\stepcounter{@auth}\relax
    \ifnum\value{@auth}=\value{@inst}\lastand\else\unskip,\ \fi}%
  \begin{center}%
  \let\newline\\
  {\Large\bfseries\boldmath\pretolerance=10000 \@title\par}\vskip.8cm
  \setbox0=\vbox{\setcounter{@auth}{1}\def\and{\stepcounter{@auth}}%
    \def\thanks##1{}\@author}%
  \global\value{@inst}=\value{@auth}%
  \setcounter{@auth}{1}%
  {\lineskip.5em\noindent\ignorespaces\@author\vskip.35cm}%
  {\small\institutename}%
  \end{center}}

\renewcommand\section{\@startsection{section}{1}{\z@}%
  {-18\p@ \@plus -4\p@ \@minus -4\p@}%
  {12\p@ \@plus 4\p@ \@minus 4\p@}%
  {\normalfont\large\bfseries\boldmath\rightskip=\z@ \@plus 8em\pretolerance=10000}}
\renewcommand\subsection{\@startsection{subsection}{2}{\z@}%
  {-18\p@ \@plus -4\p@ \@minus -4\p@}%
  {8\p@ \@plus 4\p@ \@minus 4\p@}%
  {\normalfont\normalsize\bfseries\boldmath\rightskip=\z@ \@plus 8em\pretolerance=10000}}
\renewcommand\subsubsection{\@startsection{subsubsection}{3}{\z@}%
  {-18\p@ \@plus -4\p@ \@minus -4\p@}%
  {-0.5em \@plus -0.22em \@minus -0.1em}%
  {\normalfont\normalsize\bfseries\boldmath}}
\renewcommand\paragraph{\@startsection{paragraph}{4}{\z@}%
  {-12\p@ \@plus -4\p@ \@minus -4\p@}%
  {-0.5em \@plus -0.22em \@minus -0.1em}%
  {\normalfont\normalsize\itshape}}

\renewenvironment{abstract}{%
  \list{}{\advance\topsep by0.35cm\relax\small
    \leftmargin=1cm\labelwidth=\z@\listparindent=\z@
    \itemindent\listparindent\rightmargin\leftmargin}%
  \item[\hskip\labelsep\bfseries Abstract.]}
  {\endlist}

\theoremstyle{plain}
\newtheorem{definition}{Definition}
\theoremstyle{remark}
\newtheorem{example}{Example}

\renewenvironment{thebibliography}[1]
     {\section*{\refname}
      \def\@biblabel##1{##1.}
      \small
      \list{\@biblabel{\@arabic\c@enumiv}}%
           {\settowidth\labelwidth{\@biblabel{#1}}%
            \leftmargin\labelwidth
            \advance\leftmargin\labelsep
            \itemsep\z@
            \parsep\z@
            \usecounter{enumiv}%
            \let\p@enumiv\@empty
            \renewcommand\theenumiv{\@arabic\c@enumiv}}%
      \renewcommand\newblock{\hskip .11em \@plus.33em \@minus.07em}%
      \sloppy\clubpenalty4000\widowpenalty4000%
      \sfcode`\.=\@m}
     {\def\@noitemerr
       {\@latex@warning{Empty `thebibliography' environment}}%
      \endlist}
\makeatother

\title{\theTitleBroken}
\author
{
{\miko} \inst{1}\!
\and {\will} \inst{3}\!
\and {\jpms} \inst{1,2}\!
\and {\ed} \inst{3}\!\!%
    \color{white}\thanks{
This work is partially supported by FCT grants ATTEST (CMU-PT/\-ELE/\-0009/\-2009)
and
 POLARIS (PTDC/EIA-CCO/123051/2010), by SFI grant BEACON (09/IN.1/I2618),
     and by
     Semiconductor Research Corporation  contract 2005TJ1366.
    }
}

\institute{
IST/INESC-ID, Lisbon, Portugal
\and
University College Dublin, Ireland
\and
Carnegie Mellon University, Pittsburgh, PA, USA\vspace{-0.75ex}
}

\begin{document}
\emergencystretch=3em
\maketitle
\begin {abstract}
We propose two novel approaches for using Counterexample-Guided Abstraction
Refinement (CEGAR) in Quantified Boolean Formula (QBF) solvers.
The first approach develops a recursive algorithm whose search is driven by CEGAR (rather than by DPLL). 
The second approach employs CEGAR as an additional learning technique
in an existing DPLL-based QBF solver. 
Experimental evaluation of the implemented prototypes shows that the
CEGAR-driven solver outperforms existing solvers on a number of
families in the QBF-LIB and that the DPLL solver benefits from the
additional type of learning.
Thus this article opens two promising avenues in QBF:
CEGAR-driven solvers as an alternative to existing approaches and
a novel type of learning in DPLL.%
 \end {abstract}

\section{Introduction}
Quantified Boolean formulas~(QBFs)~\cite{DBLP:series/faia/BuningB09}
naturally extend the SAT problem by enabling
expressing PSPACE-complete problems, which can be found in a number
of areas~\cite{DBLP:series/faia/GiunchigliaMN09}.
While nonrandom SAT solving has been dominated by  the DPLL procedure,
 it has proven to be far from a silver bullet
for QBF solving.  Indeed, a number of solving techniques have been proposed for QBF%
~\cite{giunchiglia2010system,DBLP:conf/lpar/Benedetti04,DBLP:conf/sat/Biere04,DBLP:conf/sat/LonsingB08,DBLP:conf/aaai/GoultiaevaB10},
complemented by a variety of {\em preprocessing techniques}~\cite{%
DBLP:conf/sat/BubeckB07,%
DBLP:conf/sat/GiunchigliaMN10,%
DBLP:conf/sat/LonsingB11,%
DBLP:conf/cade/BiereLS11}.

This paper extends the family of QBF solving techniques by employing
the counterexample guided abstraction refinement (CEGAR)
paradigm~\cite{DBLP:journals/jacm/ClarkeGJLV03}. 
This is done in two different ways. The first approach develops a
novel algorithm, named \rareqs, that gradually {\em
  expands} the given formula into a propositional one.  In contrast to
the existing expansion-based
solvers~\cite{DBLP:conf/fmcad/AyariB02,DBLP:conf/sat/Biere04,DBLP:conf/sat/LonsingB08}, 
the use of CEGAR in \rareqs enables terminating before the formula is fully expanded and
thus substantially mitigates the problems with memory blowup inherent to expansion-based solvers.
The second approach employs CEGAR as an {\em additional learning
  technique} in an existing DPLL-based QBF solver. At the price of
higher memory consumption, this learning technique enables more
aggressive pruning of the search space than the existing
techniques~\cite{DBLP:conf/iccad/ZhangM02}.
The experimental evaluation carried out demonstrates that CEGAR-based
techniques are useful for a large number of families in the QBF-LIB~\cite{QBF-library}.

\newcommand{\notation}[1]{
    \vspace{1ex}
    \noindent
    \textbf{Notation.} #1
    \vspace{1ex}
}

\section{Preliminaries} \label{sec:prelim}

Quantified Boolean formulas (QBF) are assumed, unless noted otherwise, to be in {\em prenex} form
$Q_1{z_1}{\dots}Q_n{z_n}.\phi$ where $Q_i\in\{\forall,\exists\}$, $z_i$ are distinct variables, 
and $\phi$ is a propositional formula using only the variables $z_i$ and the constants $0$ (false), $1$ (true).
The sequence of quantifiers in a QBF is called the {\em prefix} and the propositional formula the {\em matrix}.
The prefix is divided into \textit{quantifier blocks}, each of which
is a subsequence $\forall x_1\dots\forall x_n$ or resp.\ $\exists x_1\dots\exists x_n$,
which we denote by $\forall X$ or resp.\ $\exists X$, where $X = \{x_1, ..., x_n\}$.

\notation{
    We write $\bar{Q}$ for $\qt{\forall}$ (if 
    $Q$ is  $\qt{\exists}$) or
    $\qt{\exists}$ (if $Q$ is  $\qt{\forall}$).
}

Whenever convenient, parts of a prefix are denoted as~$P$ with possible subscripts, e.g.\ $P_1 \forall X P_2.\,\phi$
denotes a QBF with the matrix~$\phi$ and a prefix that contains~$\forall X$.
If the quantifier of a block $Y$ occurs within the scope of the quantifier
of another block $X$, we say that variables in $X$ are \textit{upstream} of
variables in $Y$ and that variables in $Y$ are \textit{downstream} of variables in $X$.

{\em Variable assignments} %
are represented as sets of literals.  In particular, an
assignment~$\tau$ to the set of variables~$X$ contains exactly one of
$x$, $\lnot x$ for each $x\in X$, with the meaning that if $x\in\tau$,
the variable~$x$ has the value $1$ in $\tau$ and if $\lnot x\in\tau$,
it has the value~$0$.  

\notation{
    We write $\bset^Y$ for the set of assignments to the variables~$Y$.
}

For a Boolean formula $\phi$ and an assignment $\tau$ we write
$\phi[\tau]$ for the substitution of $\tau$ in $\phi$. 
In practice a substitution also performs basic simplifications, e.g.\ 
$(\lnot x\lor y)[\{\lnot x\}]=(\lnot 0\lor y)=1$. 
We extend the notion of substitution to QBF 
so that it first removes the quantifiers of substituted variables and then substitutes all occurrences with their assigned values. E.g., if $\tau$ is an assignment to a block $X$, then
 \mathematics{\left(P_1 Q X P_2.\ \phi\right)[\tau]} results in \mathematics{P_1 P_2.\ \phi[\tau]}.

A Boolean formula in {\em conjunctive normal form (CNF)} is a
conjunction of {\em clauses},  where a clause is a disjunction of {\em
  literals}, and a literal is either a variable or its complement. 
Whenever convenient, a CNF formula is treated as a set of clauses. 
For a literal $l$, $\var(l)$ denotes the variable in $l$, i.e.\ $\var(\lnot x)=\var(x)= x$.

The pseudocode throughout the paper uses the function $\SAT(\phi)$ to
represent a call to a SAT solver on a propositional formula~$\phi$.
The function returns a satisfying assignment for $\phi$, if such
exists, and returns $\nul$ otherwise. 

\subsection {Game-Centric View}

A QBF can be seen as a
a {\em game}
between the {\em universal player} and the {\em existential player}.  During the game, the
existential player assigns values to the existentially quantified variables 
and the universal player assigns values to the universally quantified ones.
A player can assign a value to a variable only if all variables
upstream of it already have a value.  The existential player wins if
the formula evaluates to~$1$ and the universal player wins if it
evaluates to~$0$.

We note that the order in which values are given to variables in the
same block is unimportant.  Hence, by a {\em move} we mean an
assignment to variables in a certain block. A concept useful
throughout the paper are the {\em winning moves}.

\begin{definition}[winning move]\label{definition:WinningMove}
  Consider a (nonprenex) closed QBF $Q X. \Phi$ and an assignment $\tau$ to
  $X$.  Then $\tau$ is called a {\em winning move for $QX. \Phi$} if  $Q{=}\exists$ and
  $\Phi[\tau]$ is true or  $Q{=}\forall$ and $\Phi[\tau]$ is false.

\noindent\textbf{Notation.} %
   We write $\M(Q X.\,\Phi)$  to denote the set of winning moves for $Q X.\,\Phi$.
\end{definition}

\begin{observation}
  Let $\Phi$ be a QBF. 

A closed QBF $\exists X.\,\Phi$ is true iff there exists a winning move for $\exists X.\,\Phi$.

A closed QBF $\forall Y.\,\Phi$ is true iff there does {\em not} exist a winning move for $\forall Y.\,\Phi$.
\end{observation}

\section {Recursive CEGAR-based Algorithm}\label{section:recursive}

Previous work on QBF shows how CEGAR can be used to solve formulas
with 2~levels of quantifiers~\cite{JanotaSilva-SAT11}. Here we
generalize this approach to an arbitrary number of quantifiers by
recursion.
The recursion follows the prefix of the given formula starting with
the most upstream variables progressing towards more downstream
variables.
It tries to find a winning move (\autoref{definition:WinningMove}) for
variables in a certain block by making recursive calls to obtain
winning moves for the downstream variables. The base case of the
recursion, i.e., a QBF with one quantifier, is handled by a SAT
solver.

 The algorithm is presented as a recursive function returning a winning
move for the given formula, if such move exists.  Following the CEGAR
paradigm, the function  builds an abstraction which
provides {\em candidates} for the winning move. This abstraction is
gradually refined as the algorithm progresses.
Refinement is realized by {\em strengthening} the abstraction, which
means reducing the set of winning moves;
strengthening is achieved by applying conjunction and disjunction.

\begin {observation}\label{observation:expansion}
Let $\Phi_1,\dots,\Phi_n$ be QBFs with free variables in $X$.

$\M\left(\forall X.\ (\Phi_1 \lor \dots \lor \Phi_n)\right) \subseteq \M\left(\forall X.\ \Phi_i\right)$,
$i\in 1..n$.

$\M\left(\exists X.\ (\Phi_1 \land \dots \land \Phi_n)\right) \subseteq \M\left(\exists X.\ \Phi_i\right)$,
$i\in 1..n$.

$\M(\forall X\exists Y.\ \Phi) = \M(\forall X.\ \bigvee_{\mu\in\bset^Y} \Phi[\mu])$

$\M(\exists X\forall Q Y.\ \Phi) = \M(\exists X.\ \bigwedge_{\mu\in\bset^Y} \Phi[\mu])$
\end {observation}

The second half of the above observation gives us a recipe how to
eliminate quantifiers by {\em expanding} them into the corresponding
propositional operator.  One could thus eliminate quantifiers one by
one and eventually call a SAT solver if only one quantifier is
left.  The clear disadvantage of
this approach is that the formula grows rapidly and therefore performing the
expansion is often unfeasible.  This is where CEGAR comes in; the
algorithm expands quantifiers {\em carefully}, based on
counterexamples that show that the current expansion is too weak. In
this spirit, we define abstraction as a partial expansion of the given
formula.

\begin {definition}[$\omega$-abstraction]\label{definition:omega_abstraction}
Let $\omega$ be a subset of $\bset^Y$.

The $\omega$-abstraction of a closed QBF $\forall X\exists Y.\ \Phi$ is the formula $\forall X.\ \bigvee_{\mu\in\omega} \Phi[\mu]$.

The $\omega$-abstraction of a closed QBF~$\exists  X\forall Y.\ \Phi$ is the formula $\exists X.\ \bigwedge_{\mu\in\omega} \Phi[\mu]$.
\end {definition}

Observe that any winning move for~\mathematics{Q X\bar Q Y.\ \Phi} is
also a winning move for its $\omega$-abstraction (for
arbitrary~$\omega$). The reverse, however, does not hold.  Hence,
following the CEGAR paradigm, we first find a winning move for the
abstraction and then {\em verify} that it is also a winning move for
the given formula.  Verifying that a given assignment is a
winning move entails solving another QBF.

\begin{observation}\label{observation:opponent}
An assignment $\tau$ is a winning move for a closed $Q X\bar Q Y.\ \Phi$ iff
\mathematics {
\bar Q Y.\ \Phi[\tau]
}
has no winning move.
\end {observation}

If a winning move for the abstraction is verified to be a winning move
for the given formula, the move is returned. However, if this is not the
case, the abstraction is strengthened. \autoref{observation:opponent}
tells us that if an assignment $\tau$ is {\em not} a winning move for
$Q X\bar Q Y.\ \Phi$, then there {\em is} a winning move~$\mu$ for the opposing quantifier $\bar Q$ for the
QBF \mathematics{\bar Q Y.\ \Phi[\tau]}.  We say that this move~$\mu$ is a {\em
  counterexample} to $\tau$ because it serves as a witness
demonstrating that $\tau$ is not a winning move for \mathematics{Q
  X\bar Q Y.\ \Phi}.  In accordance with the concept of counterexample
{\em guided} abstraction refinement, if a counterexample $\mu$ is
found, the current $\omega$-abstraction is strengthened by adding
$\mu$ to $\omega$.

\begin{algorithm}[t]
\caption{Basic recursive CEGAR algorithm for QBF}\label{algorithm:basic_recursion}
\SetKw{Func}{Function}
\Func \solve($Q X.\,\Phi$)\;
\DontPrintSemicolon%
\Input{$Q X.\ \Phi$ is a closed QBF in prenex form with no adjacent blocks with the same quantifier}
\Output{a winning move for $Q X.\,\Phi$ if there is one, $\nul$ otherwise}
\BlankLine
\Begin{
 \If{$\Phi$ has no quantifiers}{
 \Return $(Q=\exists)\ ?\ \SAT(\phi)\ :\ \SAT(\lnot\phi)$
 }

$\omega\gets \emptyset$ \label{step:initialization}\;
\While{\true} {
$\alpha\gets (Q=\exists)\ ?\ \bigwedge_{\mu\in\omega} \Phi[\mu]\ :\ \bigvee_{\mu\in\omega} \Phi[\mu]$\tcp*[r]{build abstraction}
$\tau'\gets\solve(\prenex(Q X.\ \alpha))$\label{step:candidate}\tcp*[r]{find a candidate solution}
\lIf{$\tau' = \nul$} {%
    \Return $\nul$
}\tcp*[r]{no winning move}
$\tau\gets\comprehension{l}{l\in\tau' \land \var(l)\in X}$\label{step:filter}\tcp*[r]{filter a move for $X$}
$\mu \gets \solve(\Phi[\tau])$\label{step:counterexample}\tcp*[r]{find a counterexample}
\lIf{$\mu =\nul$}{%
    \Return $\tau$
} \;
$\omega\gets\omega\cup\{\mu\}$\label{step:refine}\tcp*[r]{refine}
}}
\end{algorithm}\DecMargin{1em}
 
When we put these things together, we obtain
\autoref{algorithm:basic_recursion}. The algorithm is given a closed QBF $Q X.\,\Phi$.
and returns a winning move for~$Q X.\,\Phi$, if such exists, and returns $\nul$ otherwise.
It is required that %
$Q X.\,\Phi$
is in prenex form where no two adjacent blocks have the
same quantifiers (the blocks are maximal).
The algorithm starts with
$\omega = \emptyset$; this represents an abstraction that can be won by any candidate.
In each iteration of the CEGAR loop it first solves the abstraction~(\autoref{step:candidate})
and then verifies whether the move winning the abstraction is also a
winning move for the given problem (\autoref{step:counterexample}).
These operations are realized as
recursive calls.  If there is no winning move for the abstraction,
then there is no winning move for the given problem and the function
terminates.  If there is no counterexample to the move winning the
abstraction, then this move is also a winning move for the given
problem and the function terminates.  If there is a counterexample to
the move winning the abstraction, the abstraction must be refined~(\autoref{step:refine}).

 The precondition of the function that the input formula must be in prenex form with
no adjacent blocks with the same quantifier poses some technical
difficulty.   When constructed directly according
to its definition (\autoref{definition:omega_abstraction}), the abstraction does not necessarily satisfy this condition. 

Consider the case
for~\mathematics{Q=\exists} (\mathematics{Q=\forall} is analogous).
 The abstraction is of the form $\exists X.\ \bigwedge_{\mu\in\omega} \Phi[\mu]$.
Prenexing the abstraction generates fresh variables for
each of the conjuncts $\Phi[\mu]$, interleaves them into a single
prefix, and merges adjacent blocks that start with the same quantifier.
Since each~$\Phi[\mu]$ starts with the existential quantifier (the substitution of $\mu$ eliminated the universal variables at
the top), after  prenexing, the
abstraction's prefix starts with $\exists X X_1 \dots X_k$ where~$X_i$
are the fresh variables for the conjuncts~$\Phi[\mu]$.  
For this reason if a winning move for the abstraction is computed, only the
assignments to the variables~$X$ are considered (\autoref{step:filter}).

\begin{example}
  Consider the QBF $\exists vw. \Phi$, where $\Phi = \forall u\exists xy.\ 
       (v\lor w\lor x)
       \land \mbox{$(\bar v\lor y)$}
       \land (\bar w \lor y)
       \land (u\lor\bar x) 
       \land (\bar u\lor\bar y)
       $,
  and the candidates 
 $\{v,w\}$ and $\{\bar v, \bar w\}$, 
  and corresponding counterexamples $\{u\}$ and $\{\bar u\}$.
  Refinement yields the  abstraction  \mbox{$\exists vw.\ \Phi[\{u\}] \land \Phi[\{\bar u\}]$}, with the prenex form
  $\exists vwxyx'y'.\ 
        (v\lor w\lor x) 
  \land (\bar v\lor y) 
  \land \mbox{$(\bar w \lor y)$}
  \land (\bar y)
  \land (v\lor w\lor x') 
  \land (\bar v\lor y') 
  \land (\bar w \lor y') 
  \land (\bar x')$
   with no winning move and the algorithm terminates with the return value $\nul$.
\end{example}

\subsection {Improving Recursive CEGAR-based Algorithm}\label{section:multigames}

 \autoref{algorithm:basic_recursion} clearly suffers from high memory consumption
since in each iteration of the loop the abstraction is increased by the size of the input formula 
 and the number of its variables is doubled (in the worst case).
Recursive calls further amplify this unfavorable behavior.
For the input formula
$\exists X.\ \Phi$, performing
 $n_1$ iterations with the counterexamples
\mathematics{\mu^1_1,\dots,\mu^1_{n_1}} yields the abstraction
\mathematics{\Omega=\exists X.\ \phi[\mu^1_1]\land\dots\land\phi[\mu^1_{n_1}]}.  
The algorithm  subsequently invokes the recursive call
$\solve(\Omega)$ on \autoref{step:candidate}.  If within this
recursive call the loop iterates~$n_2$ times, its abstraction is of the
form $\exists X.\ \Omega[\mu^2_1]\lor\dots\lor\Omega[\mu^2_{n_2}]$
with the size 
$O(n_1\times n_2 \times |\phi|)$.
In general, if the algorithm iterates~$n_i$ times at a recursion level~$i$,
the abstraction at level~$k$ is of the size
\mathematics{O(n_1\times {\dots}\times n_k\times|\phi|)}.

To cope with this inefficiency, we exploit the form of the
formulas that the algorithm handles.  In the case of the existential
quantifier, the abstraction is a conjunct, and it is a disjunct in the
case of the universal quantifier.  For the sake of uniformity, we
bridge these two forms by introducing the notion of a {\em multi-game}
where a player tries to find a move that wins multiple
formulas simultaneously.

\begin{definition}[multi-game]\label{definition:MultiGame}
 A {\em multi-game} is denoted by $Q X.\{\Phi_1,\dots,\Phi_n\}$ where
 each $\Phi_i$ is a prenex QBF starting with $\bar Q$ or has no quantifiers.
 The free variables of each $\Phi_i$ must be in~$X$ and all $\Phi_i$ have
 the same number of quantifier blocks.
 We refer to the formulas~$\Phi_i$ as {\em subgames} and $Q X$ as the
 {\em top-level prefix}.

 A {\em winning move} for a multi-game is an assignment to the
 variables~$X$ such that it is a winning move for each of the
 formulas~$Q X.\ \Phi_i$.
\end {definition}

Observe that the set of winning moves of a multi-game
$Q X.\{\Phi_1,\dots,\Phi_n\}$ is the same as the set of winning moves of the QBF
\mathematics{\forall X.(\Phi_1 \lor \dots \lor \Phi_n)}
for $Q = \forall$ %
and it is the same as \mathematics{\exists X.(\Phi_1 \land \dots \land \Phi_n)}
for \mathematics{Q=\exists}. %
And, any QBF \mathematics{Q X.\ \Phi} corresponds to a multi-game
with a single subgame \mathematics{Q X.\{\Phi\}}

To solve multi-games we use
\autoref{algorithm:multigame_recursion}. 
The algorithm is given a
multi-game to solve and the abstraction is again a multi-game.
To determine whether the candidate~$\tau$ is a winning move, it
 tests whether it is a winning move for the subgames in turn.
If it finds a subgame~$\Phi_i$ s.t.~$\Phi_i[\tau]$ is won by the
opponent~$\bar Q$ by a move~$\mu$, then $\Phi_i[\mu]$ is used to strengthen the
abstraction.

Since an abstraction is a multi-game, it seems natural to
add~$\Phi_i[\mu]$ to the set of its subgames.  This, however, cannot
be done right away because the formula is not in the right
form.  In particular, all the subgames must start with the opposite
quantifier as the top-level prefix.  Hence, if~$\Phi_i$ is of the form
\mathematics{\bar Q YQ X_1.\ \Psi_i}
and $\mu \in \bset^Y$, then 
\(\Phi_i[\mu] = Q X_1.\ \Psi_i[\mu]\).  
To bring the formula into the right form, we
introduce fresh variables for the variables~$X_1$ and move them into
the top-level prefix. More precisely, the function 
$\refine(\alpha, \Phi_l, \mu_l)$ is
defined as follows (observe that the subgames remain in prenex form).

\vspace{1pt}

\noindent\hspace{5pt}
\begin{minipage}{.95\textwidth}
$\refine \big(Q X.\{\Psi_1,\dots,\Psi_n\},\ \bar Q Y Q X_1.\,\Psi,\ \mu\big) \ :=\   Q X X_1'.\{\Psi_1,\dots,\Psi_n,  \Psi'[\mu]\}$
\end{minipage}

\begin{minipage}{.95\textwidth}
\it
 where $X_1'$ are fresh duplicates of the variables~$X_1$ and~$\Psi'$ is~$\Psi$  with~$X_1$  replaced by~$X_1'$
\end{minipage}

\noindent\hspace{5pt}
\begin{minipage}{.95\textwidth}
$\refine \big(Q X.\{\Psi_1,\dots,\Psi_n\},\ \bar Q Y.\,\psi,\ \mu\big) \ :=\  Q X.\{\Psi_1,\dots,\Psi_n, \psi[\mu]\}$
\end{minipage}

\begin{minipage}{.925\textwidth}
\it where $\psi$  is a propositional formula (where no duplicates are needed)
\end{minipage}

\vspace{3pt}

Similarly to \autoref{algorithm:basic_recursion}, after the refinement,
the abstraction's top-level prefix contains additional variables
besides the variables~$X$. Hence, values for these variables are
filtered out if a winning move for the abstraction is found.

\begin{algorithm}[t]
\caption{Recursive CEGAR algorithm for multi-games}\label{algorithm:multigame_recursion}
\Func \rareqs($Q X.\left\{\Phi_1,\dots,\Phi_n \right\}$) \;
\IncMargin{1em} %
{\bf output:}
  {a winning move for
  $Q X.\left\{\Phi_1,\dots,\Phi_n \right\}$
  if there is one; $\nul$ otherwise}
\BlankLine
\Begin{
\uIf{$\Phi_i$ have no quantifiers}{
\Return $Q=\exists\ ?\ \SAT(\bigwedge_i \Phi_i)\ :\ \SAT(\lnot(\bigvee_i \Phi))$
}
$\alpha\gets Q X.\ \{\}$ \label{step:multi_initialization}\;
\While{\true} {
$\tau' \gets\rareqs(\alpha)$%
\label{step:multi_candidate}%
\tcp*[r]{find a candidate solution}
\lIf{$\tau' = \nul$} {%
    \Return $\nul$
} \;
$\tau\gets\comprehension{l}{l\in\tau' \land \var(l)\in X}$\label{step:multi_filter}\tcp*[r]{filter a move for $X$}
\label{step:multi_counterexample}
\lFor{$i\leftarrow 1$ \KwTo $n$}{
  $\mu_i \gets \rareqs(\Phi_i[\tau])$\tcp*[r]{find a counterexample}
}
\lIf{$\mu_i =\nul$ for all $i\in\{1..n\}$}{%
    \Return $\tau$
} \; 
let $l\in\{1..n\}$ be s.t.\ $\mu_l\neq\nul$ \\
$\alpha\gets \refine(\alpha, \Phi_l, \mu_l)$
\label{step:multi_refine}\tcp*[r]{refine}
}}
\end{algorithm}\DecMargin{1em}
\subsection{Properties of the Algorithms}

In CEGAR loop of \autoref{algorithm:basic_recursion} no candidate or
counterexample repeats.  Intuitively, this is because once a
counterexample~$\mu$ is found, the abstraction is strengthened so that
in the future winning moves for the abstraction cannot be beaten by
the move~$\mu$.  Consequently, the loop is terminating and for a
formula \mathematics{QX\bar Q Y. \Phi} the number of its iterations is
bounded by the number of possible assignments to the variables~$X$
and~$Y$, i.e.~$\textsf{min}(2^{|X|}, 2^{|Y|})$.
In the worst case, in each iteration the abstraction grows by the size of $\Phi$.
For a multi-game \mathematics{Q X.\ \{ \Phi_1,\dots,\Phi_n\}} in the CEGAR loop of \autoref{algorithm:multigame_recursion} 
no candidates repeat but counterexamples may. However, for a given~$i\in 1..n$, a counterexample $\mu_i$ does not repeat. 
More precisely there are no two distinct iterations of the loop with the corresponding candidates and counterexamples %
$\tau_1,\mu_1$, $\tau_2,\mu_2$,  
such that $\mu_1=\mu_2$ and $\mu_1$ is a winning move for both $\Phi_i[\tau_1]$ and  $\Phi_i[\tau_2]$ for some $i$.
This demonstrates termination with the upper bound for the number of iterations as~$\textsf{min}(2^{|X|}, n \times 2^{|Y|})$.
In the worst case, in each iteration the abstraction grows by the maximum of the sizes of the subgames $\Phi_1,\dots,\Phi_n$.
Soundness and completeness of the algorithms~\ref{algorithm:basic_recursion} and~\ref{algorithm:multigame_recursion} are direct consequences
of \autoref{observation:expansion}.  

\subsection{Implementation Details}
We have implemented a prototype of \rareqs in {\tt C++}, supporting the
QDIMACS  format, with the underlying SAT solver {\tt
  minisat~2.2}~\cite{DBLP:conf/sat/EenS03}.
The implementation has several distinctive features.  In
\autoref{algorithm:multigame_recursion}, an abstraction computed
within a sub-call is forgotten once the call returns. This may lead to
repetition of work and hence the solver supports maintaining these
abstractions and strengthening them gradually, similarly to the way
SAT solvers provide {\em incremental} interface.  This incremental
approach, however, tends to lead to unwieldy memory consumption and
therefore, it is used only when the given multigame's subgames have 2
or fewer quantification blocks.

If an assignment $\tau$ is a candidate for a winning move that turns
out {\em not} to be a winning move, the refinement guarantees that
$\tau$ is not a solution to the abstraction in the future iterations
of the CEGAR loop. This knowledge enables us to make the subcall for
solving the abstraction more efficient by explicitly disabling~$\tau$ as
a winning move for the abstraction. We refer to this technique as {\em
  blocking} and it is similar to the refinement used in certain SMT
solvers~\cite{DBLP:conf/cade/MouraRS02,Barrett:2002}.

Throughout its course, the algorithm may produce a large number of new
formulas, either by substitution or refinement.  Since these formulas
tend to be simpler than the given one, they can be further simplified
by standard QBF {\em preprocessing} techniques. The implementation
uses unit propagation and {\em monotone} (pure) literal
rule~\cite{cadoli1998algorithm}.  These simplifications introduce the
complication that in a multi-game $Q X.\{\Phi_1,\dots,\Phi_n\}$ the
individual subgames might not necessarily have the same number of
quantifier levels. In such case, all games with no quantifiers are
immediately put into the abstraction before the loop starts.

\newsavebox{\fmbox}
\newenvironment{fmpage}[1]
{\begin{lrbox}{\fmbox}\begin{minipage}{#1}}
{\end{minipage}\end{lrbox}\fbox{\usebox{\fmbox}}}

\def\ttt{\texttt}
\def\tit{\textit}
\def\trm{\textrm}
\def\tbf{\textbf}
\def\tsf{\textsf}
\def\ss{\hspace{0.05em}}
\def\negss{\hspace{-0.05em}}
\def\ds{\displaystyle}

\newcommand{\tup}[1]{\langle #1 \rangle}
\newcommand{\itc}[1]{\textit{\it #1\/}}

\def\ttob{\texttt{\symbol{`\{}}}
\def\ttcb{\texttt{\symbol{`\}}}}
\newcommand{\ttc}[1]{\ttob{#1}\ttcb}

\def\itemb{\item[$\bullet$]}

\def\SeqSep{\models}
\newcommand{\SeqLoseRaw}[2]{#1 \SeqSep {#2 \trm{ loses}}}
\newcommand{\SeqLose}[2]{\{#1\} \SeqSep {#2 \trm{ loses}}}
\newcommand{\SeqWinRaw}[2]{#1 \SeqSep {#2 \trm{ wins}}}
\newcommand{\SeqWin}[2]{\mbox{\(\{#1\} \SeqSep {#2 \trm{ wins}}\)}}
\newcommand{\sem}[1]{\llbracket \,#1\, \rrbracket}
\newcommand{\bigsem}[1]{\big\llbracket \,#1\, \big\rrbracket}

\def\pred{\tit{pred\/}}
\def\succs{\tit{succs\/}}
\def\qb{\tit{qb}}
\def\bnfsep{\;\;|\;\;}

\section{CEGAR as a learning technique in DPLL}
The previous section shows that CEGAR can give rise to a complete and
sound algorithm for QBF. In this section we show that CEGAR enables us
to extend existing DPLL solvers with an additional learning technique.
To illustrate the basic idea consider the QBF $\forall X.\, (\exists Y.\ \phi)$
and a situation when the solver assigned values to variables in $X$ and
$Y$ such that $\phi$ is satisfied, i.e., the existential player won.
This assignment has two disjoint parts, $\pi_{\rm cand}$ and $\pi_{\rm cex}$,
which are assignments to $X$ and $Y$, respectively.
Conceptually, $\pi_{\rm cand}$ corresponds the candidate assignment in \rareqs
and $\pi_{\rm cex}$ to its counterexample.
In this case, the CEGAR-based learning will correspond to disjoining the
formula $\phi[\pi_{\rm cex}]$ onto $\phi$, resulting in 
$\forall X.\, (\exists Y.\ \phi) \lor \phi[\pi_{\rm cex}]$,
so that $\pi_{\rm cand}$ is avoided in the future.

\def\True{\tsf{T}}
\def\False{\tsf{F}}
\def\SetTrue{\{\tsf{T}\}}
\def\SetFalse{\{\tsf{F}\}}
\def\SetUnknown{\varnothing}
\def\SetError{\{\tsf{F},\tsf{T}\}}

\def\phiin{\Phi_\trm{in}}
\def\picex{\pi_c}

\def\restrict{\tit{Reduce}_{\,\picex}}
\def\Extend{\tit{Refine\/}_{\picex}}

\def\CurAsgn{\(\pi_\trm{cur}\)}
\def\TargFmla{\(\Phiin\)}
\def\InFmla{\(\phiin\)}

The CEGAR learning in DPLL is most naturally described in the context
of a non-prenex, non-clausal solver such as GhostQ
\cite{DBLP:conf/sat/KlieberSGC10}.  Given an assignment $\pi$, such a solver
will tell us that either 
(1) the existential player has a winning stategy under $\pi$ (i.e., $\phiin[\pi]$ is true),
(2) the universal player has a winning stategy under $\pi$ (i.e., $\phiin[\pi]$ is false), or
(3) it is not yet known which player has a winning strategy under $\pi$.

{
\DecMargin{-1em}
\setlength{\algomargin}{0.0em}
\begin{algorithm}[t]
\caption{DPLL Algorithm with CEGAR Learning}\label{fig:dpll-algo}
\vspace{1ex}
\begin{alltt}
\tbf{\bf \small \phantom{0}1.}   global \CurAsgn{} = \(\varnothing\); \vspace{0.5ex}
\tbf{\bf \small \phantom{0}2.}   function dpll_solve(\InFmla) \ttob
\tbf{\bf \small \phantom{0}3.}     while (true) \ttob
\tbf{\bf \small \phantom{0}4.}        while (\textsf{we don't know who has a winning strategy under \CurAsgn}) \ttc{
\tbf{\bf \small \phantom{0}5.}          decide\_lit(); propagate();
\tbf{\bf \small \phantom{0}7.}        }
\tbf{\bf \small \phantom{0}8.}        \InFmla := dpll_learn(\InFmla);
\tbf{\bf \small \phantom{0}9.}        if (\textsf{we learned who has a winning strategy under \(\varnothing\)}) return;\!\!\!
\tbf{\bf \small 10.}\textcolor{red}{        if (\tsf{last decision literal is owned by winner}) \ttob}
\tbf{\bf \small 11.}\textcolor{red}{          \InFmla := cegar\_learn(\InFmla);}
\tbf{\bf \small 12.}\textcolor{red}{        \ttcb}
\tbf{\bf \small 13.}        backtrack();
\tbf{\bf \small 14.}        propagate(); /\!/ \textsf{Learned information will force a literal.}
\tbf{\bf \small 15.}     \ttcb
\tbf{\bf \small 16.}   \ttcb
\end{alltt}
\end{algorithm}
}

\begin{figure}[tbh!]
    \begin{fmpage}{0.978\textwidth}
    \vspace{1ex}
    \begin{enumerate}
    \item Let $X_c$ be the quantifier block of the last decision literal.
    \\Let $Q_c$ and $\Phi_c$ be such that $(Q_c X_c. \Phi_c)$ is a subformula of $\phiin$.
    \item \label{dceg-cex} Let $\picex$ be a complete assignment for $X_c$ created by extending the
    solver's current assignment with arbitrary values for the unassigned variables
    in $X_c$ and removing variables in blocks other than $X_c$.
    This assignment $\picex$ corresponds to the \emph{counterexample} in the recursive CEGAR 
    approach.  
    \item \label{dceg-subst}
    We modify $\phiin$ by:
        \begin{itemize}
        \itemb
        $
        \trm{substituting } (\exists X_c. \Phi_c) \trm{ with } 
            (\exists X_c. \Phi_c) \lor  \Phi_c[\picex]
            \trm{, if } Q_c = \qt{\exists}$, or
        \itemb
        $\trm{substituting } (\forall X_c. \Phi_c) \trm{ with } 
            (\forall X_c. \Phi_c) \land  \Phi_c[\picex]
            \trm{, if } Q_c = \qt{\forall}.
        $
        \end{itemize}
    \item All variables that are bound by a quantifier inside $\Phi_c[\picex]$ are renamed
    to preserve uniqueness of variable names.
    \vspace{1ex}
    \end{enumerate}
    \end{fmpage}
    \caption{CEGAR Learning in DPLL}  \label{fig:cegar-dpll}
\end{figure}

We modify such a solver by inserting a call to a new CEGAR-learning procedure
after performing standard DPLL learning, as shown in Algorithm~\ref{fig:dpll-algo}.
We write ``$\phiin$'' to denote the current input formula, i.e., the input
formula enhanced with what the solver has learned up to now.  Both standard DPLL
learning and CEGAR learning are performed by modifying $\phiin$.  As shown in
Algorithm~\ref{fig:dpll-algo}, CEGAR learning is performed only if the last
decision literal is owned by the winner.  
(The case where the last decision literal is owned by the losing player
corresponds to the conflicts that take place {\em within} the underlying SAT solver in \rareqs.)
The CEGAR-learning procedure is shown in Figure~\ref{fig:cegar-dpll}.  
Step~\ref{dceg-subst} is justified by Observation~\ref{obs:dpll-idempot} below,
which in turn is justified by Observation~\ref{obs:dpll-expansion}.

\begin{observation}\label{obs:dpll-expansion}\rm Consider an arbitrary QBF
$(Q_c X_c .\, \Phi_c)$, possibly containing free variables, but where
each bound variable is bound by at most one quantifier.  Then it follows
immediately from definition of quantification that:
\vspace{0.33ex} \\
$$
\exists X_c.\Phi_c \;=   \bigvee_{\pi \in {\bset}^{X_c}}\hspace{-0.2em}\Phi_c[\pi]
\quad\quad
\trm{and}
\quad\quad
\forall X_c.\Phi_c \;= \bigwedge_{\pi \in {\bset}^{X_c}}\hspace{-0.2em}\Phi_c[\pi]
$$ 
\vspace{0.50ex} \\
(Recall that ``\,$\bset^{X_c}$\,'' denotes the set of all assignments to $X_c$.)
\end{observation}

\begin{observation}\label{obs:dpll-idempot}\rm Since conjunction and
disjunction are idempotent, %
\vspace{-0.5ex}
\begin{align*}
\textstyle
\exists X_c.\Phi_c = (\exists X_c.\Phi_c) \lor  \Phi_c[\pi_c], \trm{ where } \pi_c \in \bset^{X_c}
\\[-0.5ex]
\textstyle
\forall X_c.\Phi_c = (\forall X_c.\Phi_c) \land \Phi_c[\pi_c], \trm{ where } \pi_c \in \bset^{X_c}
\end{align*}
\end{observation}

\subsection{Implementation Details} \label{sec:dpll-implementation}
\def\gout{g_\trm{out}}
We have implemented a limited version of CEGAR learning in the solver GhostQ
\cite{DBLP:conf/sat/KlieberSGC10}.  
Our implementation uses a modified version of step \ref{dceg-subst} of
Figure~\ref{fig:cegar-dpll}.  We substitute $\picex$ into the original version
of the input formula $\phiin$, not the current version of $\phiin$.  
Although substituting into the original formula instead of the current formula
potentially reduces the effectiveness of CEGAR learning (since we can't learn a
refinement of a refinement), it reduces the memory consumed per refinement.
Unit propagation and the Pure Literal Rule are applied to simplify the result
of the substitution, among other optimizations.

Step~\ref{dceg-cex} of Figure~\ref{fig:cegar-dpll} extends the
counterexample~$\picex$ to a complete assignment to the quantifier block~$X_c$.
This allows completely eliminating a quantifier block, which may
cause two quantifier blocks of the same quantification type to become
adjacent to each other. If so, the two adjacent blocks are merged together,
providing greater freedom in selecting variable order.  
\section{Experimental Results}\label{section:results}
Our objective was to analyze the effect of CEGAR on the different
families of available benchmarks. Due to do the large number of
families in QBF-LIB~\cite{QBF-library}, we have targeted families
from {\it formal verification} and {\it planning} as  two prominent
applications of QBF.
Several large and hard families were sampled with $150$~files ({\tt
  terminator}, {\tt tipfixpoint}, {\tt Strategic Companies}); the area
of planning contains four classes for robot planning, each counting
1000 instances with similar characteristics and thus only one
of these classes was selected ({\tt Robots2D}).
The solvers \qube, \quantor, and \nenofex were chosen for comparison.
  \qube is a state-of-the-art DPLL-based solver;
 \quantor and \nenofex are
expansion-based solvers (c.f.\ \autoref{section:related_work}).
The experimental results were obtained on an Intel Xeon 5160 3GHz,
with 4GB of memory. The time limit was set to $800$\,seconds and  the
memory limit to 2GB.

All the instances were preprocessed by the preprocessor
{\bloqqer}~\cite{DBLP:conf/cade/BiereLS11} and instances 
solved by the preprocessor alone were excluded from further analysis.
An exception was made for the family {\tt Debug} where  preprocessing
turned out to be infeasible and the family was considered in its unpreprocessed
form.

Unlike the other solvers, \GQ's input format is not clause-based (QDIMACS) but it is
 circuit-based. To enable running \GQ on the targeted instances, the solver
was prepended with a reverse-engineering front-end.  Since this
front-end cannot handle \bloqqer's output, \GQ was run directly on the
instances without preprocessing. 
The other solvers were run on the preprocessed instances (further preprocessing was disabled for \qube).

\begin{figure}[t]%
\centering
  \includegraphics[height=.99\textwidth,angle=270]{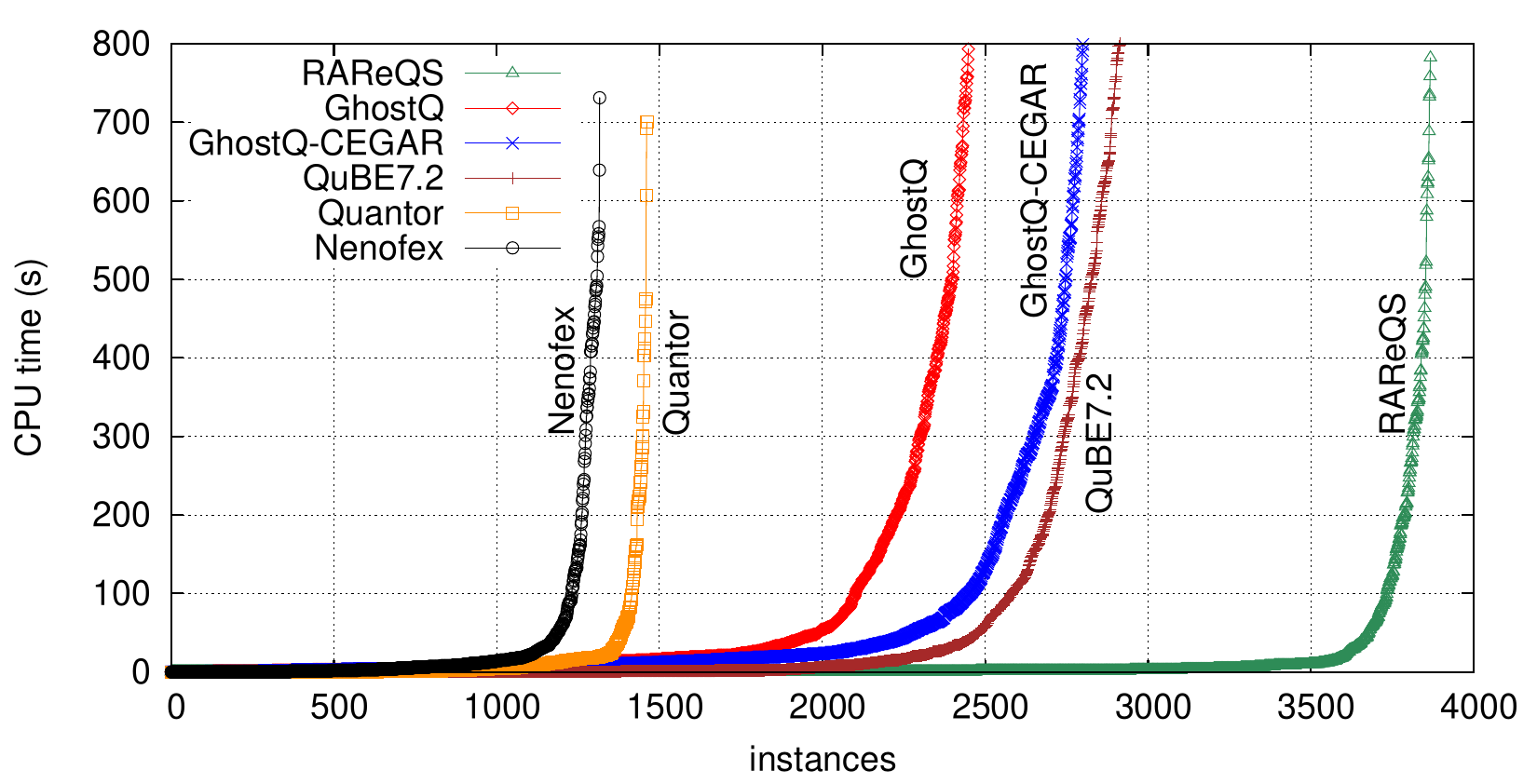}
  \caption{Cactus plot of the overall results}\label{figure:cactus}
\end{figure}

The relation between solving times and instances is presented by a cactus plot in \autoref{figure:cactus}; 
number of solved instances per family are shown in \autoref{figure:families};
 a comparison of \rareqs with other solvers is presented in \autoref{figure:differences}.
More detailed information can be found at 
 \url{http://sat.inesc-id.pt/~mikolas/sat12}.

On the considered benchmarks, \rareqs solved the most instances,
approximately $33\%$ more than the second solver \qube.  \rareqs also
turned out to be the best solver for most of the types of the
considered instances.  \autoref{figure:differences} further shows that
for each of the other solvers, there is only a small portion of
instances that the other solver can solve and \rareqs cannot.
Out of the $801$ instances when the
 solver was aborted, only~$50$ ran out of of memory.%

In several families the addition of CEGAR learning to \GQ worsened its
performance. With the exception of {\tt Robots2D}, however, the
performance was worse only slightly.  
Overall, \GQ benefited from the additional CEGAR learning and in
particular for certain families.  A family worth noting
is {\tt irqlkeapclte}, where no instances were solved by any of the
solvers except for \GQc.

The usefulness of CEGAR was in particular demonstrated by the families
{\tt incrementer-encoder}, {\tt conformant-planning}, {\tt trafficlight-controller}, {\tt Sorting-networks}, and
{\tt BMC}
where \rareqs solved significantly more instances than the existing
solvers, and \GQc improved significantly over \GQ.  Most
notably, for {\tt incrementer-encoder\;(484)} and {\tt RobotsD2\;(700)}
only one instance was not solved by \rareqs, and for {\tt
  blackbox-01X-QBF\;(320)} and {\tt trafficlight-controller\;(1459)}
\rareqs solved {\em all} instances.

\begin{table}[t]%
\centering
    {
    \def\strut{\rule[-1ex]{0pt}{3.5ex}}
    \noindent
    \begin{small}     
    \begin{tabular}{|@{\;\;}l@{\;\;}|@{\;\;}r@{\;\;}|@{\;\;}r@{\;\;}|@{\;\;}r@{\;\;}|@{\;\;}r@{\;\;}|@{\;\;}r@{\;\;}|}
    \hline
    \strut{}                & \GQ  & \GQc & \qube & \quantor  & \nenofex  \\ \hline
    \strut{}Only \rareqs    & 1661 & 1336 &  998 & 2436 & 2564 \\ \hline
    \strut{}Only competitor &  242 &  269 &   46 &   30 &   13 \\ \hline
    \end{tabular}
    \end{small}
    }

     \vspace{1.5ex}
    \caption{Number of instances solved by \rareqs but not by a competing solver, and \textit{vice versa}}
    \label{figure:differences}
\end{table}

\begin{table}[tbh!]
\centering
\begin{scriptsize}
\setlength{\arrayrulewidth}{0.25pt}
\def\strut{\rule[0.50ex]{0pt}{1.65ex}}
\setlength{\tabcolsep}{0.5em}
\noindent
\begin{tabular}{|l@{\hspace{-0.00em}}r|r|r|r|r|r|r|}
\hline
Family \strut& Lev. & {\sf RAReQS} & \GQ  & \textsf{\!\scalebox{0.9}[1.0]{GhostQ-}Cegar\!}&  \qube &  \quantor &  \nenofex  \\
\hline
\hline
trafficlight-ctlr (1459)      \strut& 1--287 & {\bf 1459}  &  {806}  &  {1001}  &  {1092}  &  {955}  &  {863} \\\hline
RobotsD2 (700)                      \strut& 2--2   & {\bf 699}  &  {350}  &  {271}  &  {630}  &  {0}  &  {30} \\\hline
incrementer-encoder (484)           \strut& 3--119 & {\bf 483}  &  {285}  &  {477}  &  {284}  &  {51}  &  {27} \\\hline
blackbox-01X-QBF (320)              \strut& 2--21  & {\bf 320}  &  {138}  &  {126}  &  {224}  &  {3}  &  {4} \\\hline
Strat.\ Comp. (samp.) (150)\!\!    \strut& 1--2   & {\bf 107}  &  {12}  &  {12}  &  {\bf 107}  &  {18}  &  {12} \\\hline
BMC (85)                            \strut& 1--3   & {\bf 73}  &  {26}  &  {48}  &  {37}  &  {65}  &  {64} \\\hline
Sorting-networks (84)               \strut& 1--3   & {\bf 72}  &  {24}  &  {32}  &  {45}  &  {38}  &  {38} \\\hline
blackbox-design (27)                \strut& 5--9   & {\bf 27}  &  {\bf 27}  &  {\bf 27}  &  {18}  &  {0}  &  {0} \\\hline
conformant-planning (23)            \strut& 1--3   & {\bf 17}  &  {7}  &  {16}  &  {5}  &  {13}  &  {12} \\\hline
Adder (28)                          \strut& 3--7   & {\bf 11}  &  {2}  &  {2}  &  {4}  &  {5}  &  {9} \\\hline
Lin.\ Bitvec.\ Rank.\ Fun.\ (60)\!\!\strut& 3--3   & {\bf 9}  &  {0}  &  {0}  &  {0}  &  {0}  &  {0} \\\hline
Ling (8)                            \strut& 1--3   & {\bf 8}  &  {6}  &  {\bf 8}  &  {\bf 8}  &  {\bf 8}  &  {\bf 8} \\\hline
Blocks (7)                          \strut& 3--3   & {\bf 7}  &  {6}  &  {\bf 7}  &  {5}  &  {\bf 7}  &  {\bf 7} \\\hline
fpu (6)                             \strut& 1--3   & {\bf 6}  &  {0}  &  {0}  &  {\bf 6}  &  {\bf 6}  &  {\bf 6} \\\hline
RankingFunctions (4)                \strut& 2--2   & {\bf 3}  &  {0}  &  {0}  &  {\bf 3}  &  {0}  &  {0} \\\hline
Logn (2)                            \strut& 3--3   & {\bf 2}  &  {\bf 2}  &  {\bf 2}  &  {\bf 2}  &  {\bf 2}  &  {\bf 2} \\\hline
Mneimneh-Sakallah (163)             \strut& 1--3   & {110}  &  {\bf 148}  &  {141}  &  {89}  &  {3}  &  {22} \\\hline
tipfixpoint-sample (150)            \strut& 1--3   & {26}  &  {\bf 128}  &  {127}  &  {22}  &  {5}  &  {6} \\\hline
terminator-sample (150)             \strut& 2--2   & {98}  &  {\bf 109}  &  {103}  &  {9}  &  {25}  &  {0} \\\hline
tipdiam (121)                       \strut& 1--3   & {55}  &  {\bf 99}  &  {93}  &  {54}  &  {21}  &  {14} \\\hline
Scholl-Becker (55)                  \strut& 1--29  & {37}  &  {\bf 43}  &  {40}  &  {29}  &  {32}  &  {27} \\\hline
evader-pursuer (15)                 \strut& 5--19  & {10}  &  {\bf 11}  &  {8}  &  {\bf 11}  &  {2}  &  {2} \\\hline
uclid (3)                           \strut& 4--6   & {0}  &  {\bf 2}  &  {\bf 2}  &  {0}  &  {0}  &  {0} \\\hline
toilet-all (136)                    \strut& 1--1   & {134}  &  {133}  &  {131}  &  {131}  &  {\bf 135}  &  {133} \\\hline
Counter (58)                        \strut& 1--125 & {30}  &  {14}  &  {11}  &  {20}  &  {\bf 33}  &  {15} \\\hline
Debug (38)                          \strut& 3--5   & {3}  &  {0}  &  {0}  &  {0}  &  {\bf 24}  &  {6} \\\hline
circuits (63)                       \strut& 1--3   & {8}  &  {4}  &  {5}  &  {5}  &  {\bf 9}  &  {8} \\\hline
Gent-Rowley (205)                   \strut& 7--81  & {52}  &  {67}  &  {67}  &  {\bf 70}  &  {2}  &  {0} \\\hline
jmc-quant (+squaring) (20)          \strut& 3--9   & {2}  &  {0}  &  {0}  &  {\bf 6}  &  {0}  &  {2} \\\hline
irqlkeapclte (45)                   \strut& 2--2   & {0}  &  {0}  &  {\bf 44}  &  {0}  &  {0}  &  {0} \\\hline
\hline
\small{total (4669)  \rule[-0.80ex]{0pt}{3.20ex}}& & {\small \bf 3868}  &  \small{2449}  &  \small{2801}  &  \small{2916}  &  \small{1462}  &  \small{1317} \\\hline
\end{tabular}
 \end{scriptsize}
    \vspace{0.7ex}
\caption{Number of instances solved within 800 seconds by each solver.  ``Lev'' indicates the number of quantifier blocks (min--max) in the family of instances, post-bloqqer.}\label{figure:families}
\end{table}

\section{Related Work}\label{section:related_work}
CEGAR has proven useful in number of areas, most notably in model
checking~\cite{DBLP:journals/jacm/ClarkeGJLV03} and SMT
solving~\cite{DBLP:conf/cade/MouraRS02,Barrett:2002}; more recently it has been applied to handle quantification in
SMT~\cite{whd10,DBLP:conf/cav/Monniaux10}.  Special cases of QBF, 
with limited number of quantifiers, have been targeted by CEGAR:
computing vertex eccentricity~\cite{DBLP:conf/sat/MneimnehS03},
nonmonotonic
reasoning~\cite{DBLP:conf/ACMse/BrowningR06,JanotaEtAl10-JELIA},
two-level quantification~\cite{JanotaSilva-SAT11}.

A SAT solver was used in~\cite{DBLP:conf/cp/SamulowitzB05} to
guide DPLL search of a QBF solver and to cut out unsatisfiable branches.
A notion of abstraction was also used in QBF
preprocessing~\cite{DBLP:conf/sat/LonsingB11}. This notion, however,
differs from the one used in \rareqs as it means treating
universally quantified variables as existentially quantified.

An important feature of \rareqs is the expansion of the given QBF into
a propositional formula, which is then solved by a SAT solver.
This technique is used for preprocessing~\cite{DBLP:conf/sat/BubeckB07}
but also several
existing solvers tackle QBF solving in this way, most notably
{\QUBOS}~\cite{DBLP:conf/fmcad/AyariB02},
\quantor~\cite{DBLP:conf/sat/Biere04}, and
\nenofex~\cite{DBLP:conf/sat/LonsingB08}.
Just as \rareqs uses multi-games, these solvers employ some various techniques  to mitigate
the blowup of the expansion (besides preprocessing). \QUBOS uses
{\em miniscoping}, {\quantor} {\em tree-like prefixes}, and \nenofex
uses {\em negation normal form}.
In these aspects, the solvers share similarities with \rareqs.

The way the expansion is carried out is significantly
different.  
While the other solvers start the expansion from the innermost variables,
\rareqs starts from the outermost variables.
The main difference, however, lies in the {\em careful expansion} in
\rareqs.  In the aforementioned solvers, once a variable is scheduled
to be expanded, both of its values are considered in the expansion%
.
In contrast, in \rareqs 
only a particular assignment to a block of variables
chosen in the expansion and the expansion is checked
whether it is sufficient or not.  This is an important factor for both
time and space complexity. For large formulas, the traditional
expansion-based solvers are bound to generate unwieldy formulas but
the use of abstraction in \rareqs enables the solver to stop before
this expansion is reached. This leads to generating easier formulas
for the underlying SAT solver and dramatically mitigates the problems with memory
blowup.
\section{Conclusions and Future Work}

Applying the CEGAR paradigm, this paper develops two novel techniques
for QBF solving.  The first technique is a CEGAR-driven solver \rareqs and the
second an additional learning technique for DPLL solvers.

In its workings, \rareqs is close to expansion-based solvers (e.g.\ \quantor, \nenofex)
but with the important difference that the expansion is done
step-by-step, driven by counterexamples.  
Thus, the solver builds an abstraction of the given formula by constructing a partial expansion.
The downside of this approach may be that if in the end a {\em full}
expansion is needed, then \rareqs performs the same expansion as a
traditional expansion-based solver but with the overhead of
intermediate tests for whether or not the expansion is already
sufficient.

However, the approach has important advantages. 
Whenever there is no winning move for the partial expansion, then there is no winning move
for the given formula. This enables \rareqs to quickly
stop for formulas with no winning moves.
For formulas for which there {\em is} a winning move, \rareqs only
needs to build a strong-enough partial expansion whose winning moves
are also likely to be winning moves for the given formula.
The experimental results demonstrate the ability of \rareqs to avoid
the inherent memory blowup of expansion solvers, and, that careful
expansion outperforms a traditional DPLL-based approach on a large number of practical instances.

We have shown that abstraction-refinement as used in \rareqs 
is also applicable within DPLL solvers as an additional learning mechanism.
This provides a more powerful learning technique than standard clause/cube
learning, although it requires more memory.
Experimental evaluation %
 indicates that this
type of learning is indeed useful for DPLL-based solvers.

In the future we plan to further develop our DPLL solver so that it supports the
full range of CEGAR learning exploited by \rareqs and to 
investigate how to fine-tune this %
learning in order to mitigate the speed penalty for the cases where
the learning provides little information over the traditional
learning. This can not only be done by better engineering of the
solver but also devising schemata that disable the learning once
deemed too costly.  In \rareqs we plan to investigate how to integrate
techniques used in other solvers. In particular, more aggressive
preprocessing as used in Quantor and techniques for finding commonalities
in formulas used in Nenofex and dependency detection~\cite{DBLP:journals/jsat/LonsingB10}.
\end{document}